\documentclass[oribibl]{llncs}%
\usepackage{amsmath}
\usepackage{amssymb}
\usepackage{amsfonts}
\usepackage{url}
\usepackage{graphicx}

\usepackage{listings}
\lstdefinelanguage{prism}{morecomment=[l]{//},keywords={AF,AG,EU,EX,AX,AU,EF,EG,bool,real,qubit,integer,channel,of,finalstateproperty,property,had,cnot,ph,X,Y,Z,process,program,endprogram,begin,end,if,fi,do,od,var,true,false},ndkeywords={LUCKY,UNLUCKY}}
\newcommand{\qimp}{\mathbin{\sqsupset}}
\newcommand{\tEX}[1]                             {{\mathsf{EX}#1}}

\newcommand{\tAF}[1]                             {{\mathsf{AF}#1}}

\newcommand{\tEU}[2]                 {\mathsf{E}{[#1\mathsf{U}#2]}}

\providecommand{\U}[1]{\protect\rule{.1in}{.1in}}

\makeatletter
\def\paragraph{\@startsection{paragraph}{4}{\z@}{-6pt}{-0.5em plus -.22em minus -0.1em}{\normalsize\it}} 
\makeatother
\def\ket#1{\ensuremath{\left|#1\right\rangle}}

\begin{document}

\author{Simon J. Gay\inst{1}\thanks{Partially supported by the EPSRC grant EP/E00623X/1 \textit{(Semantics of Quantum Computation)} and EP/F004184/1 \textit{(Quantum Computation: Foundations, Security, Cryptography and Group Theory)}.}
\and Rajagopal Nagarajan\inst{2}\thanks{Partially supported by the EU Sixth Framework Programme (Project SecoQC: \textit{Development of a Global Network for Secure Communication based on Quantum Cryptography}) and the EPSRC grant EP/E00623X/1.} %
\and Nikolaos Papanikolaou\inst{2}$^{\star\star}$%
\institute{Department of Computing Science, University of Glasgow\\ \email{simon@dcs.gla.ac.uk} \and Department of Computer Science,
University of Warwick\\ \email{\{biju,nikos\}@dcs.warwick.ac.uk}} }

\title{\textsf{QMC}: A Model Checker for Quantum Systems \\ (Tool Paper)}
\maketitle

\section{Introduction}

The novel field of quantum computation and quantum information has been growing at a rapid rate; the study of quantum information in
particular has led to the emergence of communication and cryptographic protocols with no classical analogues. Quantum information protocols
have interesting properties which are not exhibited by their classical counterparts, but they are most distinguished for their applications
in cryptography. Notable results include the unconditional security proof~\cite{Mayers01} of quantum key distribution. This result, in
particular, is one of the reasons for the widespread interest in this field.  %
Furthermore, the implementation of quantum cryptography has been demonstrated in non-laboratory settings and is
already an important practical technology. Implementations of quantum cryptography have already been commercially launched and tested by a
number of companies including MagiQ, Id Quantique, Toshiba, and NEC. The unconditional security of quantum key distribution protocols does
not automatically imply the same degree of security for actual systems, of course; this justifies the need for systems modelling and
verification in this setting.

\enlargethispage{1cm}

The benefits of automated verification techniques are well known for classical communication protocols, especially in the cryptographic
setting. Our research programme is to apply these techniques to \emph{quantum} protocols with the expectation of gaining corresponding
benefits. Our earlier work involved applying probabilistic model--checking techniques to such protocols (see e.g. \cite{nikos-conc}; it
became clear that existing techniques are not satisfactory for analysing quantum systems. Today, while simulation tools for quantum systems
abound, to our knowledge no other authors have developed a tool directly aimed at verification. In this paper we describe just such a tool,
named \textsf{QMC} (Quantum Model Checker); it allows for automated verification of properties of quantum protocols. Properties to be
verified by \textsf{QMC} are expressed using the logic QCTL~\cite{mateus-tempo}.%
\textsf{QMC}\ enables the modelling of systems which can be expressed within the quantum \emph{stabilizer formalism}; such systems are
known to be simulable in polynomial time (viz. Gottesman--Knill theorem~\cite{gottes}). The systems expressible in this formalism are
restricted, in the sense that the set of operations which they can perform is not universal for quantum computation. Nevertheless,
stabilizers are sufficient to describe a number of systems of practical interest. We assume that the reader has an understanding of the
basics of quantum computation~\cite{nielsenchuang}.

\section{Tool Description}
\textsf{QMC} is a command--line tool, implemented in Java 5, which simulates all possible executions of a quantum protocol model, and then
checks that the model satisfies any number of specification formulae supplied by the user. We will now discuss the modelling language of
\textsf{QMC}, the simulation of models, the logic used for specifying properties, and the algorithms used for model--checking. We will
briefly discuss a concrete example, namely a model of quantum coin-flipping~\cite{bb84-orig}, the \textsf{QMC} source for which is in the
appendix, and is also available at \url{http://www.dcs.warwick.ac.uk/~nikos/downloads/case-study.qmc}.

\paragraph{Modelling Language.} Protocols are modelled using a high-level, concurrent
programming language developed especially for \textsf{QMC}.  
It is an imperative language with lightweight concurrency, typed variables including channels, and commands for allocating and manipulating
the qubits in a global quantum state. Qubit variables are pointers to the individual quantum bits (qubits) in the global state. The
language has guarded commands, which are used to express looping and nondeterministic choices.

\textsf{QMC} has a statically typed language, and variables of different types must be declared at the beginning of a protocol model
(shared variables), or at the beginning of the process in which they are used (local variables). The classical data types include integers
(type \lstinline!integer!), bits (type \lstinline!bool!), and floating--point numbers (type \lstinline!real!). There is a channel type
corresponding to each of these base types (e.g. \lstinline!channel of integer!). Qubit variables are references to qubits in \textsf{QMC}'s
internal quantum state. After declaration, these variables must be initialized with the \lstinline!newqubit! expression. A process can
apply any one of the \emph{Clifford group operators} (see~\cite{nielsenchuang}): Hadamard (\lstinline!had!), Controlled--not
(\lstinline!cnot!), Phase (\lstinline!ph!) to a given qubit, as well as perform quantum measurement with respect to the computational
basis. These are the operations allowed in the quantum stabilizer formalism.

\paragraph{Simulating {\textup{\textsf{QMC}}} programs.} As in PROMELA, any command in \textsf{QMC} is either \emph{executable} or not from the current state. \textsf{QMC}'s process scheduler interleaves executable statements from all process declarations in the input, resolving all non--deterministic choices. Quantum measurement is also treated as a source of non-determinism; a qubit in a general stabilizer state always produces one of two possible outcomes with equal probability, and we treat these equiprobable outcomes like a non--deterministic choice. Consequently, we do not analyse the probabilities of different execution paths; this is an area for future work. The scheduler ensures
finiteness of models.%
We are currently developing an operational semantics for the \textsf{QMC} language. Execution of a
\textsf{QMC} program produces a tree of execution paths, in which the branching arises from non--determinism.

\paragraph{Specification Logic.} The properties of a given protocol model are expressed using the logic QCTL~\cite{mateus-tempo}, which is a temporal logic especially
designed for reasoning about quantum systems based on CTL~\cite{clarke-emerson-sistla-CTL}. This logic allows us to reason about the
evolution of the global quantum state as a given protocol model is executed. It also enables reasoning about the states of classical
variables and, hence, measurement outcomes. 
The reader is referred to~\cite{mateus-tempo} for full details of the syntax of the logic, which includes classical formulae ($\alpha =
\bot \;|\; \textsf{qb} \;|\; \alpha \Rightarrow \alpha$), terms ($t = x \;|\; (t+t) \;|\; (t t) \;|\; \mathrm{Re}(\ket{\top}_A) \;|\;
\mathrm{Im}(\ket{\top}_A) \;|\; \smallint\phi$), quantum formulae ($\gamma = (t \leq t) \;|\; \bot \;|\; (\alpha \qimp \alpha)$) and
temporal formulae ($\theta = \gamma \;|\; \theta \qimp \theta \;|\; (\tEX \theta) \;|\; (\tEU{\theta}{\theta}) \;|\; (\tAF{\theta})$).


Interesting state formulas are those of the form~$\int\phi \leq a$ and~$[\mathsf{qb}_i,\mathsf{qb}_j]$, where~$\phi$ is a base formula
(e.g.~$\neg \mathsf{qb}_0$) and~$\mathsf{qb}_i,\mathsf{qb}_j,\mathsf{qb}_0$ are qubit variables. The first formula states that, the
probability of formula $\phi$ being satisfied in the current state is less than or equal to~$a$. The
formula~$[\mathsf{qb}_i,\mathsf{qb}_j]$ states that, in the current state, the qubits corresponding to the variables $\mathsf{qb}_i$ and
$\mathsf{qb}_j$ are \emph{not entangled} with the rest of the quantum system. In order to evaluate such a formula, \textsf{QMC} analyses
the entanglement of the current quantum state.

\paragraph{Verification Algorithms.}
\textsf{QMC} implements algorithms for evaluating EQPL formulas over stabilizer states, which are represented internally using a matrix
representation (see \cite{aaronson}). In order to check the truth of a particular formula, its truth needs to be determined for all
possible valuations; the tool automatically extracts all valuations from the internal representation. More interestingly, the tool has been
designed to explore all possible executions of a particular protocol arising from different measurement outcomes and non--deterministic
choices. Entanglement formulae may be checked without converting the internal representation to the set of all valuations. In fact, it is
possible to determine whether a list of qubits $q_i,q_{i+1},\ldots$ constitutes a partition of the global quantum state at any point during
execution using the \emph{entanglement normal forms} developed by Audenaert and Plenio~\cite{audenaert}. Temporal formulae are checked by
traversing the tree produced by the execution of a protocol model.

\paragraph{An Example.} We have built a \textsf{QMC} model for the quantum coin--flipping protocol due to Bennett and
Brassard~\cite{bb84-orig} which is available online at the URL mentioned at the beginning of Section 2. Quantum coin--flipping enables two
users, Alice and Bob, to establish a common random bit $x$ through the transmission of a single qubit $q$ and its measurement. The protocol
relies on the principle that, if Alice and Bob use compatible bases for preparation and for measurement of this qubit, their bit values
will be guaranteed to match by the laws of quantum mechanics. Incompatible bases will produce a matching bit value only with probability
$\frac{1}{2}$, although in this case the protocol is repeated and the bit discarded. There are various possible attacks that may be
performed by an enemy, which would enable him or her to compromise the final bit value. It is such attacks that we would like to
investigate using the \textsf{QMC} tool on this and related quantum
protocols. The basic property which needs to be checked at the end of the protocol is that the bit values of Alice and Bob   %
do indeed match, and this is only true if the measurement basis $\hat{b}$ chosen by Bob matches the preparation basis $b$ of Alice. Also,
if Alice's and Bob's bits and bases are in agreement, the protocol should not abort; this is expressed as a temporal property.

\section{Conclusion and Future Work}

We have described \textsf{QMC}, a model-checking tool for quantum protocols. As far as we know, it is the first dedicated verification tool
(as opposed to simulation systems) for quantum protocols. \textsf{QMC} allows the modelling and verification of properties of protocols
expressible in the quantum stabilizer formalism. The logic for expressing properties is QCTL. We have considered a simple example
illustrating the input language of the tool. It is significant to note that the restriction to quantum stabilizer states allows
\textsf{QMC} to simulate protocols efficiently, although it limits the expressive power of the tool. There are protocols which involve
quantum states that fall outside the scope of the stabilizer formalism, and we are currently investigating ways of approximating such
states using techniques due to Bravyi and Kitaev~\cite{bravyi}. Using such techniques it will be possible to obtain an implementation which
is powerful enough for the analysis of general quantum protocols and the detection of potential flaws in their design.

\enlargethispage{1cm}


\appendix
\section*{Appendix:\\ \textsf{QMC} Model of the Quantum Coin-Flipping Protocol}\enlargethispage{3cm}
\begin{lstlisting}
program QuantumCoinFlipping; var AtoB, BtoA: channel of qubit;
    A2B, B2A: channel of bool;

process Alice; var x, b, g, result: bool; q: qubit; begin
    q := newqubit;
    {Choose one of the four BB84 states and prepare qubit q}
    if
    :: true -> x:=false; b:=false;
    :: true -> x:=false; b:=true; had q;
    :: true -> x:=true; b:=false; X q;
    :: true -> x:=true; b:=true; X q; had q;
    fi
    AtoB!q;
    B2A?g;
    A2B!b;
    A2B!x;
    result := ((not (b and g)) and (b or g));
end;

process Bob; var g, x_hat, b_hat, x, b, result: bool; rq: qubit;
    abort: bool;        {Bob aborts the protocol}
    dontknow: bool;     {Bob cannot determine Alice's honesty}
begin
    AtoB?rq;
    if
    :: true -> g:=false; B2A!g;
    :: true -> g:=true; B2A!g;
    fi
    {Select random measurement basis and measure}
    if
    :: true -> b_hat:=false;
    :: true -> b_hat:=true; had rq;
    fi
    x_hat:=meas rq;
    {Receive original b and x}
    A2B?b;
    A2B?x;
    {Compare bases and bits}
    if
    :: ((b=b_hat) and (not (x=x_hat))) -> abort:=true;
    :: (not (b=b_hat)) -> dontknow:=true;
    fi
    result := ((not (b and g)) and (b or g));
end;
endprogram.

finalstateproperty (Alice.result == Bob.result);

property (AG (((b==b_hat) and (x==x_hat)) imp (abort==false)))
\end{lstlisting}
\end{document}